\documentclass[%
reprint,
 amsmath,amssymb,
 aps,
]{revtex4-1}

\usepackage{graphicx}
\usepackage{dcolumn}
\usepackage{bm}
\usepackage[mathlines]{lineno}
\usepackage[colorlinks=true,citecolor=blue,linkcolor=blue,anchorcolor=green, anchorcolor=green,urlcolor=blue]{hyperref}
\usepackage{mathrsfs}
\usepackage{xcolor}
\usepackage{float}
\usepackage{ulem}
\usepackage[utf8]{inputenc}

\begin{document}


\title{Circular orbit of a test particle and phase transition of a black hole}

\author{Ming Zhang}
\email{mingzhang@mail.bnu.edu.cn}
\altaffiliation[Also at ]{Department of Physics and Astronomy, University of Waterloo, Waterloo, Ontario N2L 3G1, Canada}
\author{Shan-Zhong Han}
\email{szh@mail.bnu.edu.cn}
\altaffiliation[Also at ]{Perimeter Institute for Theoretical Physics, Waterloo, Ontario N2L 2Y5, Canada}
\author{Jie Jiang}
\email{jiejiang@mail.bnu.edu.cn}
\author{Wen-Biao Liu}
\email{wbliu@bnu.edu.cn}
\affiliation{Department of Physics, Beijing Normal University, Beijing, 100875, China}

\date{\today}

\begin{abstract}
The radius of the circular orbit for the time-like or light-like test particle in a background of general spherically symmetric spacetime is viewed as a characterized quantity for the thermodynamic phase transition of the corresponding black hole. We generally show that the phase transition information of a black hole can be reflected by its surrounding particle's circular orbit.
\end{abstract}


\maketitle


\section{Introduction}
The discovery that a black hole possesses temperature and entropy \cite{Hawking:1974sw,Bekenstein:1973ur} provides us with thermodynamic method to study the strong gravity system. Investigations of thermodynamics for asymptotically anti-de Sitter (AdS) or asymptotically flat black hole have lasted for several decades, ever since the proposition of Hawking-Page phase tranition for the Schwarzchild-AdS black hole \cite{Hawking:1982dh}.

The phase transitions of the asymptotically AdS black holes have been studying intensively. That investigation, on some extent, could be divided into two stages. In the first stage, the negative cosmological constant of the asymptotically AdS black hole is viewed merely as a constant itself \cite{Chamblin:1999tk,Chamblin:1999hg,Caldarelli:1999xj}, which results in the correlation between the spacetime action and its thermodynamic Helholtz free energy, as well as the interpretation of the black hole's mass as its intermal energy. 

In the second stage, the negative cosmological constant of the AdS black hole is viewed as a thermodynamic variable, named as the thermodynamic pressure \cite{Kastor:2009wy}, which shows us that the mass of the black hole is in fact related to its enthalpy and the spacetime action is correlated with its thermodynamic Gibbs free energy \cite{Dolan:2010ha}. Accordingly, the recognition of the resemblance between the thermodynamics of AdS black hole and everyday thermodynamics has been raised to a new level \cite{Kubiznak:2012wp}. 

The asymptotically flat black hole can be viewed as a special case of the asymptotically AdS black hole with vanishing cosmological constant. Correspondingly, thermodynamics of the asymptotically flat black hole, which is not so complicated like the AdS counterpart, can be viewed as a zero-cosmological constant case of the asymptotically AdS one \cite{Cvetic:1999ne,Emparan:2007wm}.

Phase transition of the AdS black hole, in view of the AdS/CFT (Conformal Field Theory) correspondence \cite{witten1998ads,maldacena1998jm,Gubser:1998bc,Emparan:1999pm}, can be explained on the thermal CFT. For example, Hawking-Page phase transition can be regarded as a confinement-deconfinement phase transition in the dual quark gluon plasma \cite{Hawking:1982dh}. In the geometrothermodynamics, the phase transition of the AdS black hole can be reflected by the singularity of the thermodynamic intrinsic curvature scalar \cite{weinhold1975f,Ruppeiner:2012uc,Quevedo:2007mj,Hendi:2015rja} as well as the thermodynamic extrinsic curvature scalar \cite{Mansoori:2016jer,Zhang:2018djl,Wang:2018civ}. To investigate the phase transition of the black hole from different perspectives, including the recent interests in detecting the relation between the particle orbit around the black hole and the phase transition \cite{Wei:2017mwc,Wei:2018aqm,Bhamidipati:2018yqy}, which will be the starting point of this paper, can deepen our comprehend of the nature of the spacetime and the black hole.

The relation between the light-like circular orbit and phase transition for the Reissner-Nordstr\"{o}m-AdS (RN-AdS) black hole, along with the relation for Kerr-AdS background, was discovered that the radius of the phonton orbit, being similar to the event horizon of the black hole, can be a characterized quantity for the judgment of the phase transition for the black hole \cite{Wei:2017mwc,Wei:2018aqm}. The research was subsequently generalized to a case of the time-like particle's circular orbit in a backgroud of the RN-AdS black hole \cite{Bhamidipati:2018yqy}. In this paper, we will generally show that the radius of the circular orbit of the light-like or time-like particle around a spherically symmetric (asymptotically flat or AdS) black hole can indeed be well-behaved in reflecting the phase transition of the black hole.

In Sec. \ref{parttwo}, we will analyze the property of the circular orbit for a particle rotating around a spherically symmetric black hole. In Sec. \ref{partthree}, we will inspect the phase transition of the black hole in view of the circular orbit. In Sec. \ref{partfour}, as an example, we will exemplify our general result by analyzing the relation between the innermost stable circular orbit (ISCO) of a time-like particle and the phase transition of the RN-AdS black hole. Sec. \ref{con} will be devoted to our conclusion.

\section{Circular Orbit of a particle}\label{parttwo}
We consider a static spherically symmetric spacetime, which can be generally described by the metric
\begin{equation}\label{geo}
ds^2=f(r)dt^2-\frac{dr^2}{g(r)}-r^2 (d\theta^2 +\sin^{2}\theta d\phi^2).
\end{equation}
The surface gravity of the black hole is
\begin{equation}\label{kap}
\kappa=\frac{1}{2}\lim_{r\to r_{h}} \sqrt{\frac{g(r)}{f(r)}}\frac{df(r)}{dr},
\end{equation}
where $r_h$ is the radius of the event horizon. Then the temperature of the black hole can be defined as
\begin{equation}\label{temp}
T\equiv\frac{\kappa}{2\pi k_B},
\end{equation}
where $k_B$ is the Boltzmann constant. 

The Lagrangian of the particle's geodesic motion in the spacetime Eq. (\ref{geo}) can be written as
\begin{equation}
2\mathcal{L}=f(r)\dot{t}^2 -\frac{\dot{r}^2}{g(r)} -r^2{\dot{\theta}}^2 -(r^2 \sin^{2}\theta) \dot{\phi}^2,
\end{equation}
where a dot means a derivative with respect to the affine parameter $\lambda$ of the geodesic. Without loss of generality, we will set $\theta=\pi/2$ hereinafter. The canonical momenta related to the spherically symmetric spacetime can be derived from the Lagrangian as
\begin{equation}\label{cone}
p_t =\frac{\partial \mathcal{L}}{\partial \dot{t}} =f(r)\dot{t}=E,
\end{equation}
\begin{equation}
p_r=-\frac{\partial \mathcal{L}}{\partial \dot{r}}=\frac{\dot{r}}{g(r)},
\end{equation}
\begin{equation}\label{conl}
p_\phi =-\frac{\partial \mathcal{L}}{\partial \dot{\phi}} =r^2 \dot{\phi}=L,
\end{equation}
where $E,~L$ are conserved energy and angular momentum of the particle, respectively.
Then the Hamiltonian $\mathcal{H}$ of the particle can be written as
\begin{equation}
\mathcal{H}=p_t \dot{t}+p_r \dot{r}+p_\phi \dot{\phi}-\mathcal{L},
\end{equation}
so that we have
\begin{equation}\label{hami}
2\mathcal{H}=E\dot{t}-L\dot{\phi}-\frac{1}{g(r)}{\dot{r}}^2=\epsilon,
\end{equation}
where $\epsilon=1$ is for the time-like particle on the geodesic, and $\epsilon=0$ is for the light-like particle. Substituting Eqs. (\ref{cone})-(\ref{conl}) into Eq. (\ref{hami}), we have
\begin{equation}
\dot{r}^2=g(r) \left(\frac{E^2}{f(r)}-\frac{L^2}{r^2}-\epsilon\right).
\end{equation}
Then the effective potential $V_e$ of the system can be defined by the equation
\begin{equation}
V_e +\dot{r}^2 =0,
\end{equation}
from which the expression of the effective potential can be obtained as
\begin{equation}\label{effepo}
V_e =g(r) \left(\epsilon+\frac{L^2}{r^2}-\frac{E^2}{f(r)}\right).
\end{equation}
According to the definition, the particle can only move in the region where $V_e \leqslant 0$.

The circular orbit of the time-like or light-like particle can be yielded once we impose requirements
\begin{equation}\label{veff1}
V_e (r_i)=0,
\end{equation}
\begin{equation}\label{veff2}
V_e^\prime (r_i)=0
\end{equation}
onto the effective potential, where the prime denotes a derivative with respect to the coordinate $r$ and $r_i$ means the radius of the particle's circular orbit. The first condition Eq. (\ref{veff1}) means that the radial velocity of the particle vanishes, and the second condtion Eq. (\ref{veff2}) tells us that there is no radial acceleration for the particle. From Eq. (\ref{veff1}), we can obtain
\begin{equation}\label{funcf}
f(r_{i})=\frac{E^2 r_i^2}{L^2+\epsilon r_i^2 }.
\end{equation}
According to Eq. (\ref{veff2}), we have
\begin{equation}
g(r_i ) \left(\frac{E^2 f'(r_i )}{f(r_i )^2}-\frac{2 L^2}{r_i^3}\right)+\left(\epsilon+\frac{L^2}{r_i^2}-\frac{E^2}{f(r_i)}\right) g'(r_i )=0,
\end{equation}
from which we can obtain
\begin{eqnarray}\label{fprime}
f^{\prime}(r_{i})&=&\frac{f\left(r_i\right) g'\left(r_i\right)}{g(r)}+\frac{2 L^2 f\left(r_i\right){}^2}{E^2 r_i^3}-\frac{\epsilon f\left(r_i\right){}^2 g'\left(r_i\right)}{E^2 g\left(r_i\right)}\nonumber\\&&-\frac{L^2 f\left(r_i\right){}^2 g'\left(r_i\right)}{E^2 r_i^2 g\left(r_i\right)}.
\end{eqnarray}
Combining Eqs. (\ref{funcf}) and (\ref{fprime}), we get
\begin{equation}
f^{\prime}(r_{i})=\frac{2 E^2 L^2 r_i}{(L^2 +\epsilon r_i^2)^2},
\end{equation}
which tells us
\begin{equation}\label{con2}
f^{\prime}(r_{i})>0.
\end{equation}

\section{Phase transition of the black hole in view of a particle's circular orbit}\label{partthree}
Generally, there must be a relation between the event horizon $r_h$ of the black hole which satisfies $f(r_h)=g(r_h)=0$ and the circular orbit $r_i$ of the particle which meets the requirements Eqs. (\ref{veff1}) and (\ref{veff2}), such as 
\begin{equation}\label{req}
r_h =r_h (r_i ),~\text{or}~r_i=r_i (r_h).
\end{equation}
According to Eqs. (\ref{con2}) and (\ref{req}), we have
\begin{equation}
f^{\prime}(r_{i})= \frac{df(r_i )}{dr_i}=\frac{df(r_h )}{dr_h}\frac{dr_h}{dr_i}>0.
\end{equation}
Considering Eqs. (\ref{kap}) and (\ref{temp}), we have
\begin{equation}
\frac{df(r_h)}{dr_h}>0,
\end{equation}
because the temperature of the black hole should be positive. Then we can obtain
\begin{equation}\label{con1}
\frac{dr_h}{dr_i}>0,
\end{equation}
which implies that there is a monotonously increasing relation between the event horizon radius of the black hole and the circular orbit radius of the particle.

Though there are different phase transitions for black holes, such as Hawking-Page phase transiton \cite{Hawking:1982dh,Zhang:2018rlv}, Reentrant phase transition \cite{Altamirano:2013ane,Zou:2016sab}, we here only discuss the van der Waals (vdW)-like phase transition \cite{Kubiznak:2012wp}. In most cases, there are vdW-like phase transitions for AdS black holes with hairs such as the electric charge \cite{Gunasekaran:2012dq,Cai:2013qga,Wei:2012ui,Zhang:2014jfa,Mirza:2014xxa,Zhang:2015ova,Hennigar:2015esa,Hendi:2015xya,Kubiznak:2016qmn,Hennigar:2016xwd,Kuang:2016caz,Nam:2018sii,Mo:2018rks}. The critical point of the vdW-like phase transition for the black hole can be obtained by the equation
\begin{equation}\label{critione}
\frac{d T\left(r_h\right)}{d r_h}\Big{|}_{r_h=r_{hc}}=\frac{d^2 T\left(r_h\right)}{d r_h^2}\Big{|}_{r_h=r_{hc}}=0,
\end{equation}
where $r_{hc}$ is the critical event horizon radius of the black hole. We just want to know whether this relation could be transferred to the particle-circular-orbit-radius case. As
\begin{equation}\label{cor3}
\frac{d T\left(r_h\right)}{d r_h}\Big{|}_{r_h=r_{hc}} =\frac{dT(r_i)}{dr_i}\frac{dr_i}{dr_h}\Big{|}_{r_h=r_{hc}}=0,
\end{equation}
we can know that there must exist a critical circular orbit radius $r_{ic}$ for a particle as
\begin{equation}\label{rel}
\frac{dT(r_i)}{dr_i}\Big{|}_{r_i=r_{ic}}=0
\end{equation}
after considering Eq. (\ref{con1}). Besides, the relation
\begin{equation}
T(r_h)=T(r_h(r_i))=T(r_i)
\end{equation}
has been kept in mind. From the relation
\begin{eqnarray}
&&\frac{d^2 T\left(r_h\right)}{d r_h^2}\Big{|}_{r_h=r_{hc}}=\left[\frac{d}{dr_h}\left(\frac{d T(r_h)}{d r_h}\right)\right]\Big{|}_{r_h=r_{hc}}\nonumber\\&&=\left[\frac{d^2 T(r_i)}{d r_i^2}\frac{dr_i}{dr_h}+\frac{dT(r_i)}{dr_i}\frac{d}{dr_i}\left(\frac{dr_i}{dr_h}\right)\right]*\frac{d r_i}{d r_h}\Big{|}_{r_h=r_{hc}}\nonumber\\&&=0
\end{eqnarray}
together with the conditions Eq. (\ref{con1}) and Eq. (\ref{rel}), we can obtain
\begin{equation}\label{cor4}
\frac{dT(r_i)}{dr_i}\Big{|}_{r_i=r_{ic}}=\frac{d^2 T\left(r_i\right)}{d r_i^2}\Big{|}_{r_i=r_{ic}}=0.
\end{equation}
From Eqs. (\ref{critione}) and (\ref{cor4}), we can know that the critical point on the $T-r_h$ diagram corresponds to the one on the $T-r_i$ diagram.
Apart from this, we can also know that
\begin{equation}\label{corre1}
\frac{d T\left(r_h\right)}{d r_h}=\frac{dT(r_i)}{dr_i}\frac{dr_i}{dr_h}>0~\text{or}~<0 
\end{equation}
can correspond to 
\begin{equation}\label{corre2}
\frac{dT(r_i)}{dr_i}>0~\text{or}~<0,
\end{equation}
respectively as we have the relation Eq. (\ref{con1}), which means that the correspondence between $T-r_h$ diagram and $T-r_i$ diagram exists beyond the critical points.

\begin{figure*}[!htb] 
    \centering
    \includegraphics[width=7in]{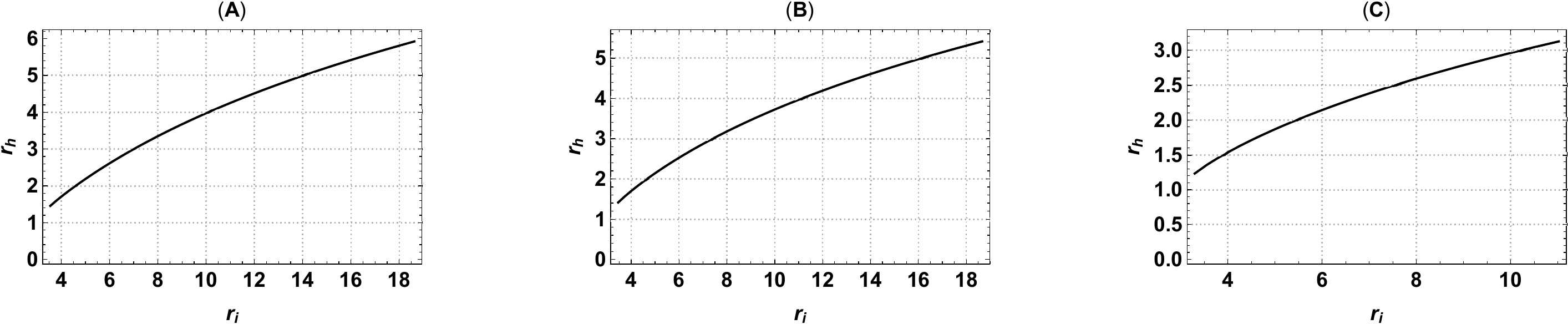}
    \caption{The event horizon $r_h$ of the black hole in terms of the radius $r_i$ of ISCO for the particle with $Q=1$ and (A) $l=\sqrt{675/(4\pi)}<l_{c}$, (B) $l=6=l_{c}$, (C) $l=5 \sqrt{3/(2\pi)}>l_{c}$.}
    \label{pic1}
 \end{figure*}

\begin{figure*}[!htb] 
    \centering
    \includegraphics[width=7in]{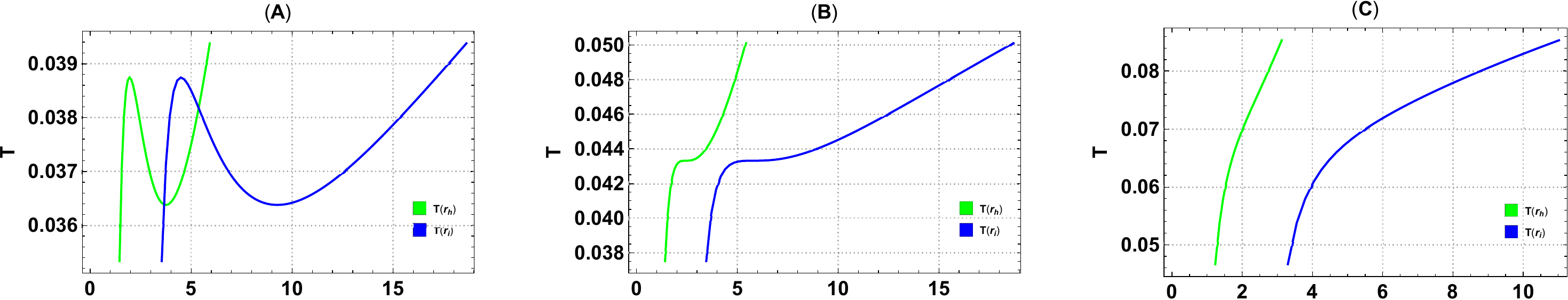}
    \caption{The temperature $T$ of the black hole in terms of the event horizon $r_h$ and the radius of ISCO $r_i$ for the particle respectively with $Q=1$ and (A) $l=\sqrt{675/(4\pi)}<l_{c}$, (B) $l=6=l_{c}$, (C) $l=5 \sqrt{3/(2\pi)}>l_{c}$.}
    \label{pic2}
 \end{figure*}

\section{An example: ISCO of a time-like particle and phase transition of the RN-A\lowercase{d}S black hole}\label{partfour}
We will show that the radius of the circular orbit for the time-like particle can be an eligible quantity to reflect the phase transition information of the black hole, using RN-AdS black hole as a representative example. Specifically, we here will focus on a special kind of the circular orbit, named as the ISCO \cite{chandrasekhar1985mathematical,wald1984general,Pugliese:2010ps,Liu:2017fjx,Pugliese:2011py,Chakraborty:2013kza,Isoyama:2014mja,Zaslavskii:2014mqa,Pugliese:2013zma,Zhang:2017nhl,Zhang:2018omr}. To obtain parameters (such as radius $r_i$) of the ISCO, we should add one another condition
\begin{equation}\label{veff3}
V^{\prime\prime}_e (r_i)=0
\end{equation}
to the effective potential Eq. (\ref{effepo}) of the particle-black hole system, which claims that the maximal and the minimal values of the effective potential merge at the ISCO radius $r_i$.

The mertric of the RN-AdS black hole can be obtained while
\begin{equation}
f(r)=g(r)=1-\frac{2 M}{r}+\frac{Q^2}{r^2}+\frac{r^2}{l^2}
\end{equation}
is given in Eq. (\ref{geo}), where $M, Q$ are respectively the mass and electric charge of the black hole, $l$ is the AdS radius of the spacetime. The event horizon $r_h$ of the black hole makes both the blackening factor $f(r)$ and $g(r)$ vanish. The mass and temperature of the black hole can be expressed in terms of the event horizon $r_h$ as
\begin{equation}\label{mass}
    M=\frac{r_h^3}{2 l^2}+\frac{r_h}{2}+\frac{Q^2}{2 r_h},
\end{equation}
\begin{equation}
    T=\frac{3 r_h}{4 \pi  l^2}+\frac{1}{4 \pi  r_h}-\frac{Q^2}{4 \pi  r_h^3}.
\end{equation}
Correspondingly, the event horizon can be got from Eq. (\ref{mass}) in a form of
\begin{equation}\label{even}
r_h=r_h (M,Q,l).
\end{equation}

In what follows, we will discuss the vdW-like phase transition for the RN-AdS black hole. In the phase transition, the critical event horizon $r_{hc}$, temperature $T_c$, AdS radius $l_c$ of the black hole which can be got from Eq. (\ref{critione}) respectively are \cite{Kubiznak:2012wp}
\begin{equation}
r_{hc}=\sqrt{6} Q,~T_c=\frac{\sqrt{5}}{18\pi Q},~l_c =6Q.
\end{equation}
When $T<T_c$, the temperature versus the event horizon of the black hole shows oscillation behaviour. When $T>T_c$, the oscillation behaviour disappears. 

According to Eq. (\ref{effepo}), the effective potential for the particle-black hole system can be written as 
\begin{equation}\label{veffrn}
V_e =\left(1+\frac{L^2}{r^2}\right)\left(1-\frac{2 M}{r}+\frac{Q^2}{r^2}+\frac{r^2}{l^2}\right)-E^2.
\end{equation}
After using the constraint conditions Eqs. (\ref{veff1}), (\ref{veff2}) and (\ref{veff3}) for the effective potential Eq. (\ref{veffrn}), one can obtain the radius $r_i$ of ISCO which can be expressed as a function of mass $M$, electric charge $Q$, and AdS radius $l$, as
\begin{equation}\label{isco}
    r_i=r_i (M,Q,l).
\end{equation}
According to Eqs. (\ref{even}) and (\ref{isco}), we can know that once we choose one of $M, Q, l$ as  parametric quantity, we can find a one-to-one mapping between the event horizon $r_h$ of the black hole and the radius $r_i$ of ISCO for the time-like particle, which can be shown as
\begin{equation}
r_h =r_h(r_i)|_{l=\text{const},Q=\text{const}}
\end{equation}
or
\begin{equation}
    r_i =r_i(r_h)|_{l=\text{const},Q=\text{const}}.
\end{equation}

By numerical calculation, we can obtain the relation, which exactly satisfies the derived result Eq. (\ref{con1}), between the event horizon radius $r_h$ and ISCO radius $r_i$, as shown in Fig. \ref{pic1}, where $r_i$ monotonously increases with $r_h$. 

In Fig. \ref{pic2}, we find the relation between the temperature $T$ and the event horizon radius $r_h$ as well as the relation between the temperature $T$ and the ISCO radius $r_i$. From the figures, we can see that the $T-r_h$ diagram and the $T-r_i$ diagram share the synchronized variation trend. The correspondence between $T'(r_h)$ and $T'(r_i)$, as well as between $T''(r_h)$ and $T''(r_i)$, showed in Sec. \ref{partthree} has been conspicously corroborated.

\section{Conclusion}\label{con}
In this paper, we find that circular orbit of the time-like or light-like particle around the spherically symmetric black hole can reflect the phase transition information of the black hole. We meticulously elaborate that there is a correspondence between the $T-r_h$ diagram and the $T-r_i$ diagram, including the critical point (in the vdW-like phase transition) and the changing tendency (which in fact reflects the thermodynamic stabibility) of the black hole. Specifically, we investigate the relation between the time-like particle's ISCO radius and the RN-AdS black hole's vdW-like phase transition, as an example. Our result is applicable to all spherically symmetric black hole, including the hairy ones \cite{Bizon:1990sr,Lavrelashvili:1992ia,Nunez:1996xv,Heisenberg:2017xda,Ganchev:2017uuo,Peng:2017gss,Dykaar:2017mba}. The result also provides a general prospect connecting particle's motion around the black hole with the phase transition of the black hole, based on which investigations of the black hole thermodynamics and phase transition are on some extent pushed forward.

\section*{Acknowledgements}
This work is supported by the National Natural Science Foundation of China (Grant Nos. 11235003, 11775022, and 11375026). We would like to thank Shao-Wen Wei for his valuable discussion.

\end{document}